\documentclass{ws-procs9x6}
\begin{document}

\title{Quark orbital angular momentum:\\
       can we learn about it from GPDs and TMDs?}

\author{H.~AVAKIAN$^1$, A.~V.~EFREMOV$^2$, \\ 
	P.~SCHWEITZER$^3$, O.~V.~TERYAEV$^2$, P.~ZAVADA$^4$}
\address{
  $^1$ Jefferson National Accelerator Facility, Newport News, VA 23606, U.S.A.\\
  $^2$ Bogoliubov Laboratory of Theoretical Physics, JINR, 141980 Dubna, Russia\\
  $^3$ Department of Physics, University of Connecticut, Storrs, CT 06269, U.S.A.\\
  $^4$ Institute of Physics, Academy of Sciences of the Czech Republic, 
       Na Slovance 2, CZ-182 21 Prague 8, Czech Republic}

\begin{abstract}
  It is known how to access information on quark orbital 
  angular momentum from generalized parton distribution functions,
  in a certain specified framework.
  It~is intuitively expected, that such information can be accessed also
  through transverse  momentum dependent distribution functions,
  but  not known how.
  Now quark models provide promising hints. 
  Recent results are reviewed.
\end{abstract}


\bodymatter

\section{Introduction}
\label{Sec:Introduction}

Transverse parton momentum dependent distribution functions (TMDs)
and generalized parton distribution functions (GPDs) describe 
complementary aspects of the transverse nucleon structure. 
TMDs \cite{Boer:1997nt,Collins:2002kn,Belitsky:2002sm} 
describe the {\sl momentum distribution} of partons in the transverse plane. 
GPDs \cite{Mueller:1998fv,Ji:1996nm,Radyushkin:1997ki,Goeke:2001tz}
describe their {\sl spatial distribution} in the transverse plane
\cite{Burkardt:2002hr}
(and much more \cite{Polyakov:2002yz}).

It is known how to learn from GPDs about orbital
angular momentum of partons in the nucleon, namely \cite{Ji:1996ek} 
(using impact parameter presentation \cite{Burkardt:2002hr})
\begin{equation}
     L^q = \int{\rm d} x \int{\rm d}^2 b\,
     \biggl(xH^q(x,b)+xE^q(x,b)-\widetilde{H}^q(x,b)\biggr)
\end{equation}
where $\int{\rm d}^2 b\,\widetilde{H}^q(x,b)$ is the helicity distribution
and $b$ the impact parameter. This decomposition has the advantage
that all spin contributions are measurable quantities.
Other decomposition schemes exist \cite{Jaffe:1989jz} and give, 
in gauge theories, in general different results \cite{Burkardt:2008ua}. 

\begin{figure}
\vspace{-8mm} \ \hspace{0.5cm}
    \psfig{file=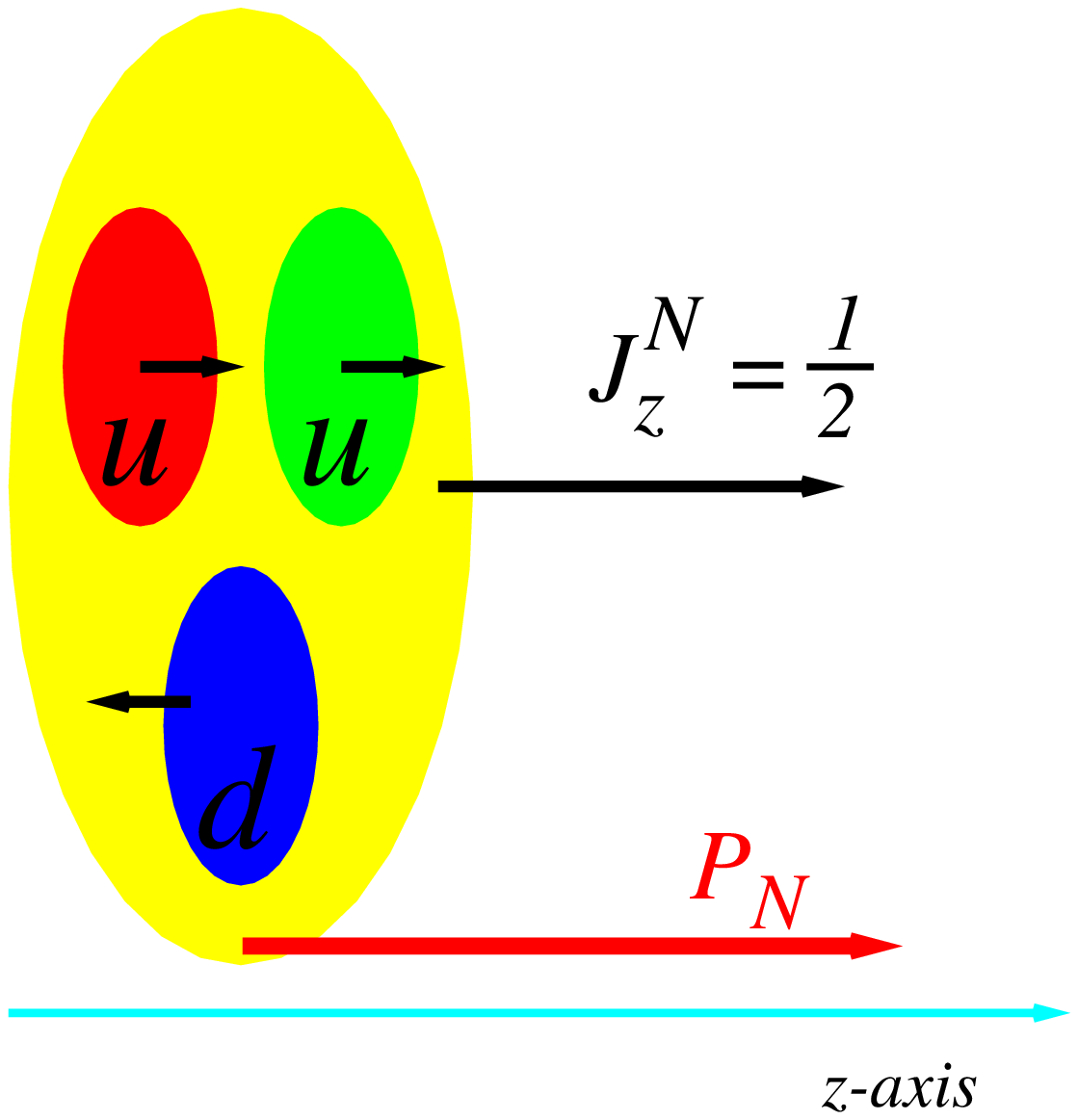,height=5cm} \ \
    \psfig{file=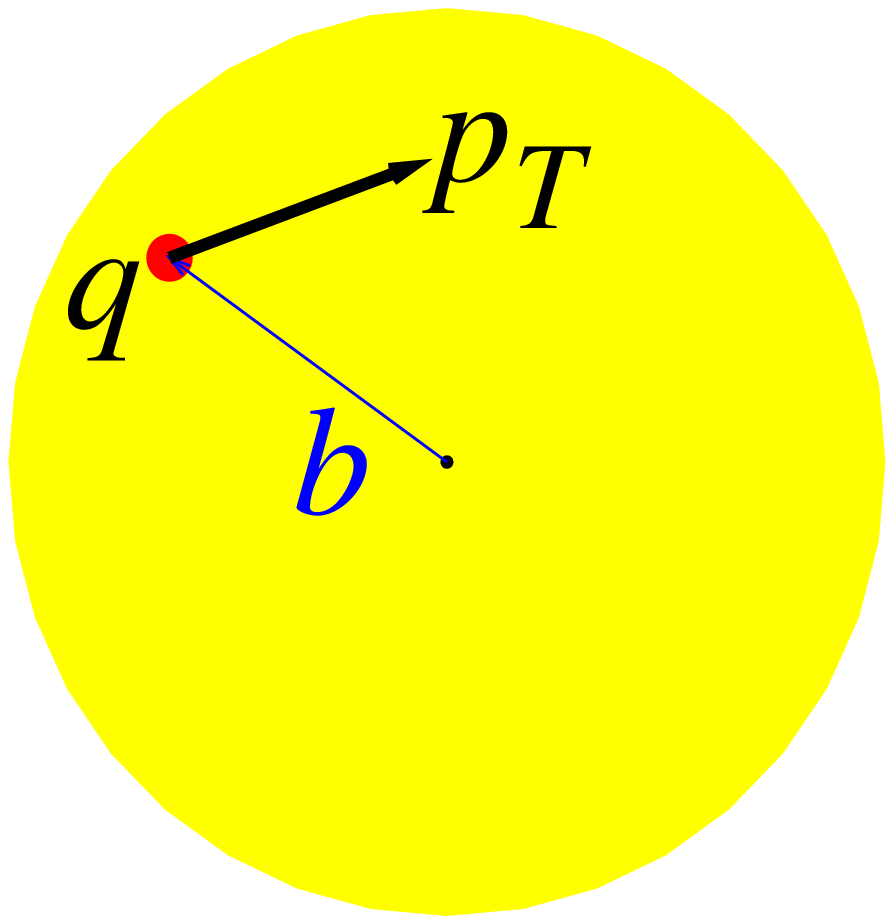,height=5.5cm}
\vspace{-2mm}
\caption{\label{Fig-1:naive-picture}
    Left: Naive picture of the spin decomposition of a
    nucleon moving along $z$-axis with the large momentum $P_N\to\infty$.
    Right: Now the nucleon moves towards us. The impact parameter
    $b$-distribution of the quark $q$ is described by GPDs.
    The complimentary information on its transverse momentum
    $p_T$-distribution is described by TMDs.}
\end{figure}

In this way, one obtains from the spatial distribution of partons 
in the transverse plane information about orbital angular momentum.
TMDs contain information on the parton momenta $p_T$ in the
transverse plane. This is in some sense complimentary to GPDs,
see Fig.~\ref{Fig-1:naive-picture}.
Intuitively, one would therefore expect TMDs to contain also
information about orbital angular momentum. However, so far
no rigorous connection of orbital angular momentum and TMDs
could be established.

Recent results from quark models could indicate a possible
connection, and the key to that is ``pretzelosity.'' 
We review the recent developments.

\section{The key TMD: pretzelosity}
\label{Sec-2:review-pretzelosity}

The light-front correlator (with \cite{Collins:2002kn,Belitsky:2002sm}
a process-dependent gauge-link ${\cal W}$)
\begin{equation}\label{Eq:correlator}
    \phi(x,\vec{p}_T)_{ij} = \int\frac{{\rm d} z^-{\rm d}^2\vec{z}_T}{(2\pi)^3}
    \;e^{ipz}\;
    \langle  N(P,S)|\bar\psi_j(0)\,{\cal W}\,\psi_i(z)|N(P,S)\rangle
    \biggl|_{{z^+=0\;\;\;\;\,}\atop{p^+ = xP^+}} \;\;
    \end{equation}
allows us to define the twist-2 chiral-odd TMDs of the nucleon as
\begin{eqnarray}\label{Eq:chiral-odd-TMDs}
    \frac12\;{\rm tr}\biggl[i\sigma^{+j}\gamma_5 \;\phi(x,\vec{p}_T)\biggr] 
    = S_T^j\,h_1  + S_L\,\frac{p_T^j}{M_N}\,h_{1L}^\perp 
    + \frac{\varepsilon^{jk}p_T^k}{M_N}\,h_1^\perp\; &&\nonumber\\
    + \;\;
    \frac{(p_T^j p_T^k-\frac12\,\vec{p}_T^{\:2}\delta^{jk})S_T^k}{M_N^2}\,
    h_{1T}^\perp \,.&& 
\end{eqnarray}
All TMDs in (\ref{Eq:chiral-odd-TMDs}) depend on $x$ and 
$p_T^{}=|\vec{p}_T^{}|$, and can be accessed in semi-inclusive deeply inelastic
scattering (SIDIS) \cite{Boer:1997nt} in combination with the Collins
fragmentation function \cite{Collins:1992kk} $H_1^\perp$ by measuring 
various azimuthal (single spin) asymmetries. In particular
\begin{equation}\label{Eq:AUTpretzel}
   A_{UT}^{\sin(3\phi-\phi_S)} \propto 
   \frac{\sum_q e_q^2 \;h_{1T}^{\perp q}\otimes H_1^{\perp q}}
        {\sum_q e_q^2 \;f_1^q \otimes D_1^q}
   \:.
\end{equation}

Positivity bounds \cite{Bacchetta:1999kz} constrain 
$|h_{1T}^{\perp(1)q}(x,p_T^{})|\le\frac12(f_1^q(x,p_T^{})-g_1^q(x,p_T^{}))$ with
the (1)-moment defined as 
$h_{1T}^{\perp(1)q}(x,p_T^{}) \equiv p_T^2/(2M^2)\,h_{1T}^{\perp q}(x,p_T^{})$.

In the limit of a large number $N_c$ of colors in QCD 
\cite{Pobylitsa:2003ty} pretzelosity behaves as 
$(h_{1T}^{\perp u}+h_{1T}^{\perp d})/(h_{1T}^{\perp u}-h_{1T}^{\perp d})\sim 1/N_c$.
Interesting aspects are \cite{Miller:2007ae,Burkardt:2007rv}
that it describes the ``non-sphericity'' of the 
spin distribution of quarks in a transversely polarized nucleon,
and requires \cite{Miller:2007ae,Burkardt:2007rv} the presence 
of nucleon wave-function components differing by two units
of orbital angular momentum. At large $x$ it is predicted 
\cite{Burkardt:2007rv,Avakian:2007xa} to behave as 
$h_{1T}^{\perp q} \sim (1-x)^5$. That is all that 
is known about this function model-independently.

What raised much interest about this TMD are results from quark models. 
The following relation was found in the bag model in 
Ref.~[\refcite{Avakian:2008dz}] 
\begin{equation}\label{Eq:pretzel-relation}
   h_{1T}^{\perp(1)q}(x,p_T^{}) = g_1^q(x,p_T^{}) - h_1^q(x,p_T^{}) \;.
\end{equation}
This relation is supported also in other
\cite{Jakob:1997wg,Meissner:2007rx,Pasquini:2008ax,Efremov:2009ze,
She:2009jq,Avakian:2010br,in-preparation,Bacchetta:2008af} 
though not all \cite{Bacchetta:2008af} 
quark models, and is broken when gauge field degrees of freedom are present
\cite{Meissner:2007rx}, see [\refcite{Avakian:2009jt}] for a review. Notice 
that $h_{1T}^{\perp(1)q}(x,p_T^{}) \equiv p_T^2/(2M^2)\,h_{1T}^{\perp q}(x,p_T^{})$, 
and if we recall that the difference of $g_1$ and $h_1$ vanishes 
in the non-relativistic limit \cite{Jaffe:1991ra}, we see that this 
(1)-moment of pretzelosity is a 'measure of relativistic effects' 
in the nucleon. This is not surprizing because in this limit
\cite{Efremov:2009ze} 
\begin{equation}
     \lim_{\rm non\mbox{-}rel}\,
     h_{1T}^{\perp q}(x,p_T^{}) =  -\frac{N_c^2}{2}\;
     P_q\;\delta\!\left(x-\frac{1}{N_c}\right) \;\delta^{(2)}(\vec{p}_T^{})
\end{equation}
where $P_u=\frac43$, $P_d=-\frac13$, and similarly for other TMDs,
i.e. in the non-relativistic limit the (1)-moments of all TMDs vanish.
Another interesting model relation is the remarkable 
{\sl non-linear relation} first observed in the covariant parton model 
\cite{Efremov:2009ze} connecting all T-even, chiral-odd, 
twist-2 TMDs:
\begin{equation}\label{Eq:rel-non-linear}
      \frac{1}{2}\,\biggl[h_{1L}^{\perp q}(x,p_T^{})\biggr]^2 
      \;\,\stackrel{\rm model}{=}\;\, 
      - \,h_1^q(x,p_T^{})\,h_{1T}^{\perp q}(x,p_T^{})\;.
\end{equation}
This relation implies an attractive prediction. The transversity
distribution gives rise to the single spin asymmetry
$A_{UT}^{\sin(\phi+\phi_S)} \propto \sum_qe_q^2\,h_1^q\otimes H_1^{\perp q}$
whose sign is known experimentally
\cite{Airapetian:2004tw}.
From (\ref{Eq:rel-non-linear}) it immediately follows that
the pretzelosity asymmetry in (\ref{Eq:AUTpretzel}) must have 
opposite sign \cite{Avakian:2009jt}, i.e.\
\begin{equation}
                    {\rm sign}\,[\,A_{UT}^{\sin(3\phi-\phi_S)}\,]
     = (-\,1) \cdot {\rm sign}\,[\,A_{UT}^{\sin( \phi+\phi_S)}\,]\;,
\end{equation}
which is expected \cite{Avakian:2009jt} to hold in the valence-$x$ 
region, where this prediction is confirmed by COMPASS for negative
hadrons from a deuteron target.\cite{Alekseev:2010dm}.

\section{Pretzelosity and Orbital Angular Momentum}
\label{Sec-3:pretzel-OAM}

Many more quark model relations among TMDs were found, 
see [\refcite{Avakian:2010br}] for the derivation of a complete set 
of relations in the bag model, and it is understood why they are 
widely supported in a large class of quark models \cite{Pasquini:2010new}.

The pretzelosity-relation (\ref{Eq:pretzel-relation}) plays a particularly
important role in what follows for the following reason. Namely, it 
was shown in the light-cone SU(6) quark-diquark model \cite{Ma:1998ar} 
that the contribution to the nucleon spin from the orbital angular 
momentum of quarks is related to the difference of transversity 
and helicity distributions, i.e.\ to the right-hand-side of the
pretzelosity-relation (\ref{Eq:pretzel-relation}) found in
\cite{Avakian:2008dz}.
It was subsequently shown that the pretzelosity-relation 
(\ref{Eq:pretzel-relation}) is valid also in the light-cone SU(6) 
quark-diquark model \cite{She:2009jq}. In other words, the (1)-moment 
of pretzelosity is, in this quark model, a measure for the 
contribution of quark orbital angular momentum to the nucleon spin.
This exciting finding was subsequently confirmed in the bag model
\cite{Avakian:2010br} and the covariant parton model 
\cite{in-preparation}.

More precisely, three different quark models 
\cite{She:2009jq,Avakian:2010br,in-preparation}
support the relation
\begin{equation}\label{Eq:OAM-pretzel}
       L_z^q =  - \int{\rm d} x\,h_{1T}^{\perp(1)q}(x)\;
\end{equation}
(in principle, there is a fourth model where it holds: in the 
non-relativistic limit one has the consistent result \cite{Efremov:2009ze}
$L_z^q =  - \int{\rm d} x\,h_{1T}^{\perp(1)q}(x)=0$).

An interesting question in this context: how can chiral-odd
(pretzelosity) and chiral-even (orbital angular momentum) 
quantities be related?

The answer in the bag model is as follows. Here
the quark wave-function has an upper-($s$-wave-)component
and a lower-($p$-wave-)component. The expectation value of 
the orbital angular momentum operator in the $s$-wave is zero,
i.e.\ only the $p$-wave contributes.
Next, we know \cite{Miller:2007ae,Burkardt:2007rv} 
pretzelosity requires $\Delta L = 2$ which in the bag model 
is possible only through interference of the $L_z=\pm 1$
components of the $p$-wave, i.e.\ again only the $p$-wave
contributes. 
Finally, knowing that only the $p$-wave (i.e.\ the lower
component of the Dirac-spinor) matters, we can ``replace'' 
in the operator of pretzelosity 
$\gamma^0$ = ${\rm diag}(1, -1)$ (in Bjorken-Drell
notation) 
by $(-1)\times$(unit matrix). This changes the number of 
gamma-matrices by one unit, and ``transforms'' a chiral-odd 
operator into a chiral-even one \cite{Avakian:2010br}.

From this exercise we learn: the relation between pretzelosity
and orbital angular momentum is at best at the level of matrix 
elements. In other words, there is no operator-identity between
these quantities --- not even in quark models.
It is interesting to stress that in quark models the result for
$L_z^q$ 
does not depend on which orbital angular momentum 
definition \cite{Jaffe:1989jz,Ji:1996ek} 
is used \cite{Burkardt:2008ua}.

\section{How does pretzelosity look like,  and how to access it?}
\label{Sec-4:pretzel-in-models}

\begin{figure}[t!]
\hspace{-1mm}\psfig{file=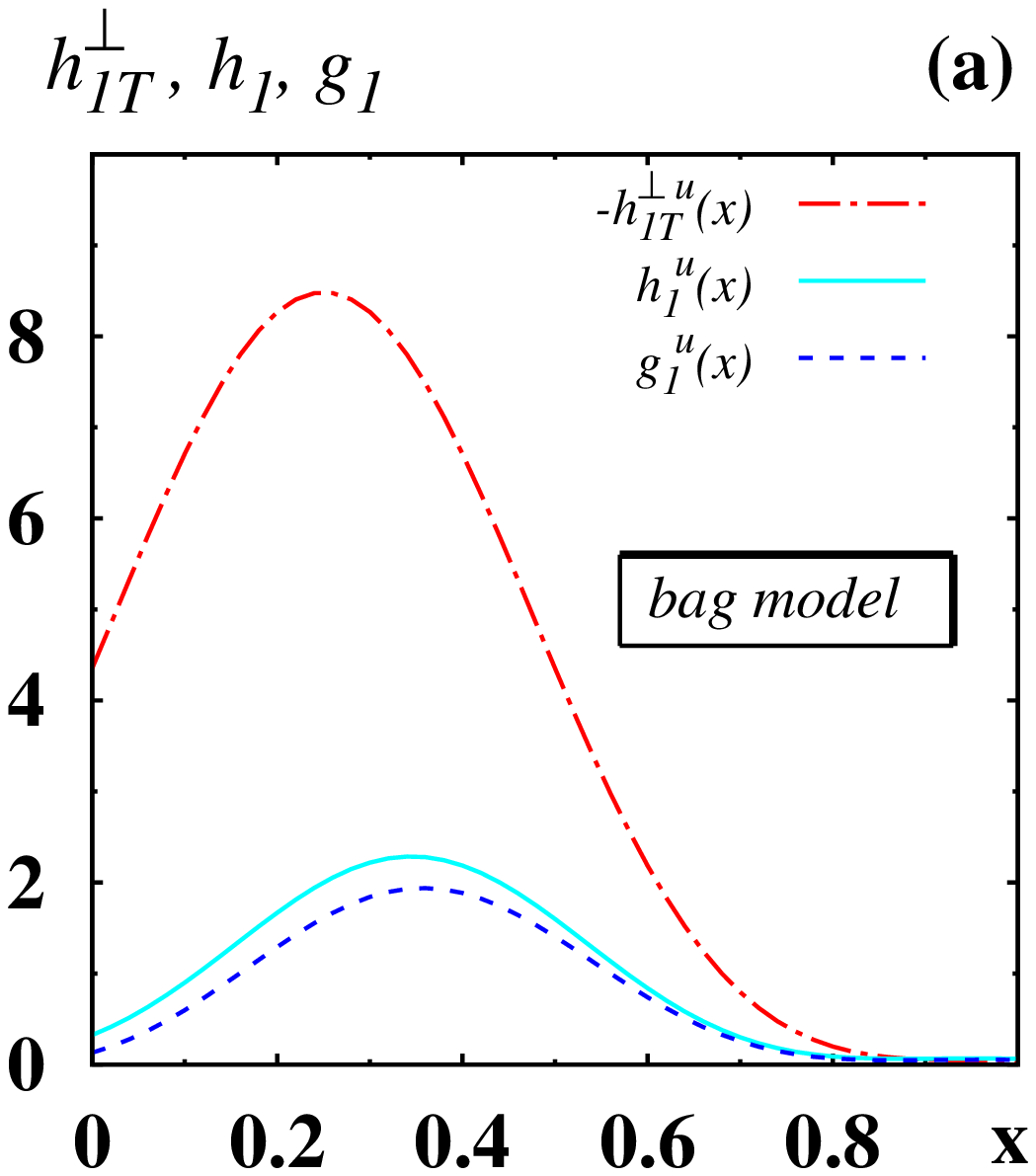,height=4.0cm}
\hspace{-4mm}\psfig{file=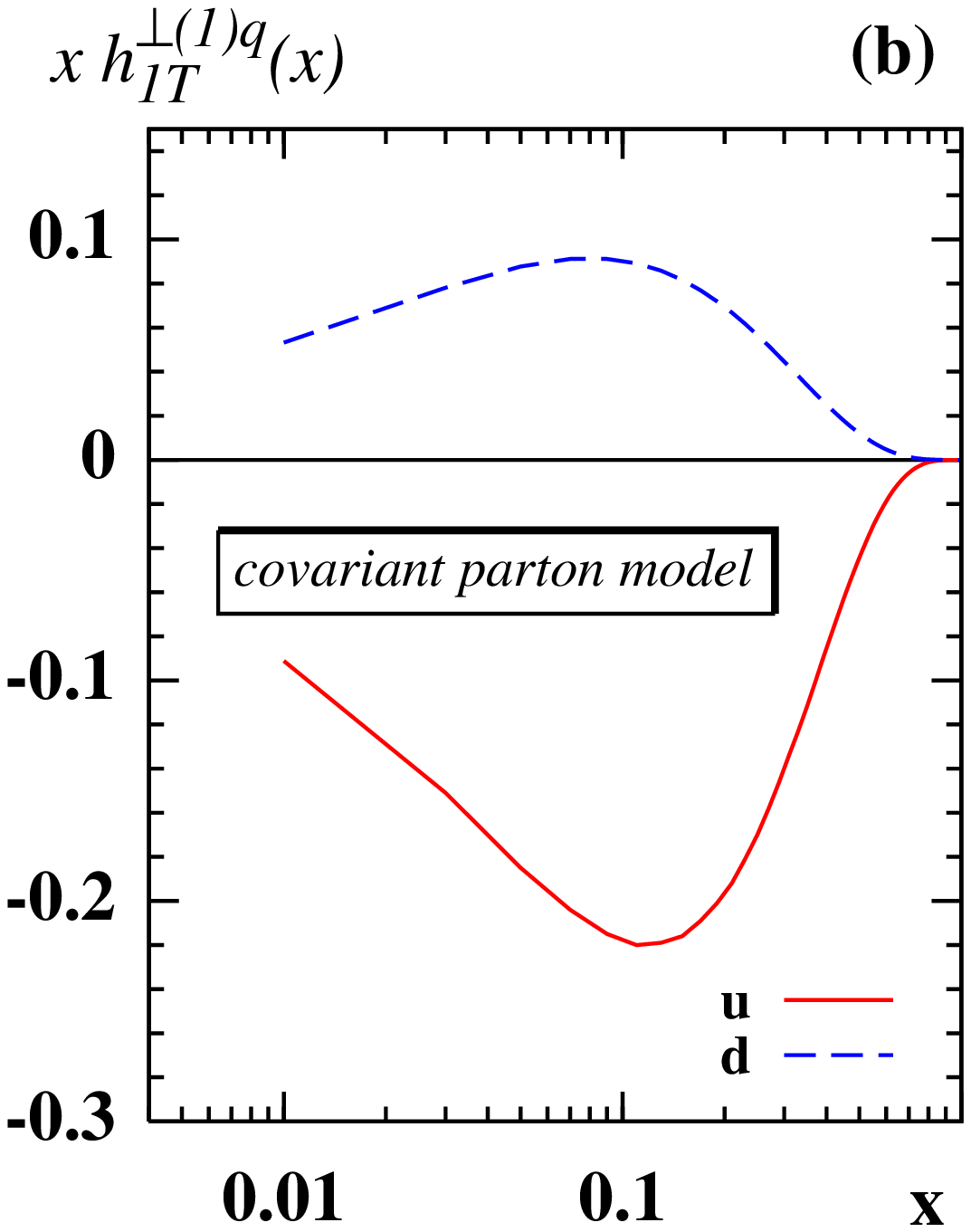,height=4.0cm}
\hspace{-1mm}\psfig{file=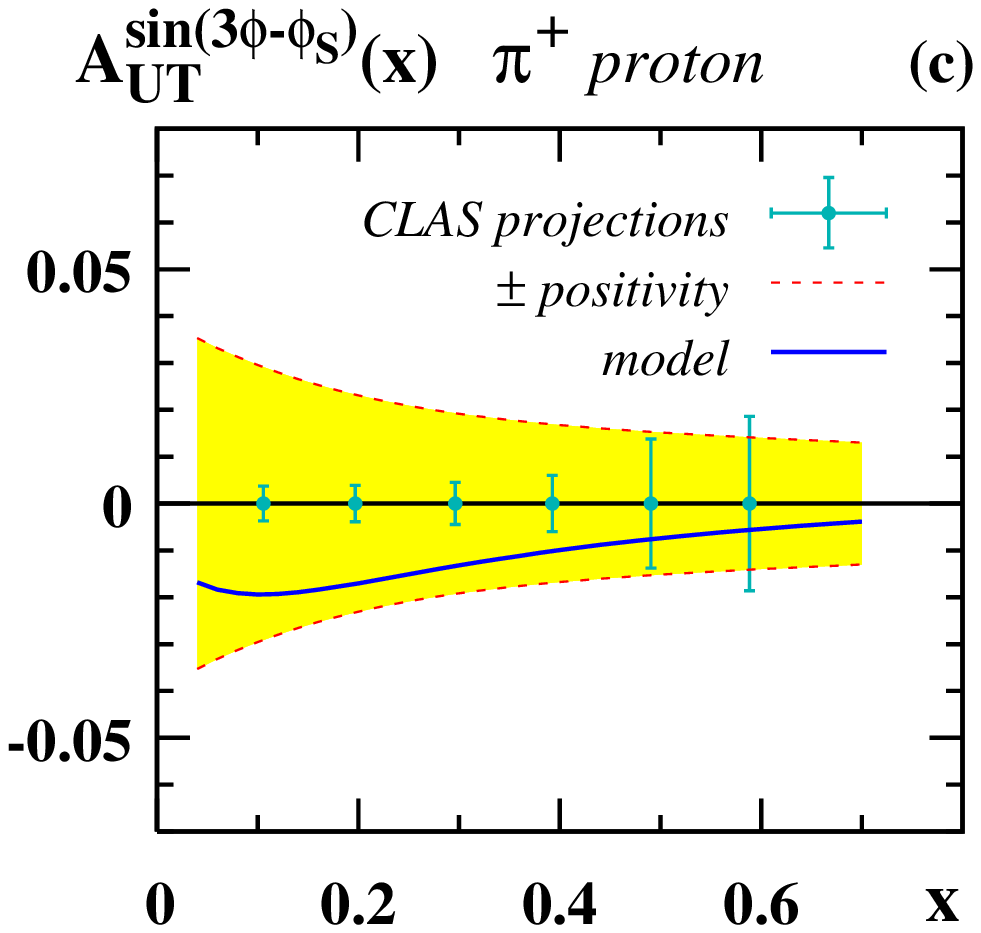,height=4.0cm}
\caption{\label{Fig-2:TMDs-model}
    (a) 
    Bag model predictions at the low scale for
    $(-1)h_{1T}^{\perp u}(x)$,  $h_1^u(x)$, $g_1^u(x)$ vs.~$x$.
    (b) 
    $h_{1T}^{\perp(1) q}(x)$ from the covariant parton model 
    \cite{Efremov:2009ze} at the high scale $Q^2 = 4\,{\rm GeV}^2$ vs.~$x$.
    (c) 
    The azimuthal single spin asymmetry $A_{UT}^{\sin(3\phi-\phi_s)}$,
    Eq.~(\ref{Eq:AUTpretzel}) for $\pi^+$ production from a proton 
    target in the kinematics of the CLAS experiment with 12 GeV 
    upgrade vs.~$x$. The shaded areas show \cite{Avakian:2008dz}
    what is allowed by positivity (quark models predict 
    negative sign). The solid curve is the prediction \cite{Efremov:2009ze} 
    from the covariant parton model shown in 
    Fig.~\ref{Fig-2:TMDs-model}b. The error projections 
    are for a 2000 hours run time \cite{Avakian-LOI-CLAS12}.}
\end{figure}

Having discussed that in quark models pretzelosity is related to orbital 
angular momentum, it is interesting to ask what quark models actually predict.
Fig.~\ref{Fig-2:TMDs-model}a shows the bag model predictions for 
$h_{1T}^{\perp u}(x)$ in comparison to $h_1^u(x)$ and $g_1^u(x)$.
The negative sign and the large magnitude of $h_{1T}^{\perp u}(x)$ 
can be understood from the non-relativistic limit \cite{Efremov:2009ze}, 
where pretzelosity is enhanced by the factor $\frac12N_c^2$ compared 
to $h_1^u(x)$ or $g_1^u(x)$, see Sec.~\ref{Sec-2:review-pretzelosity}. 
Relativistic models (like the bag model) preserve this enhancement.
Concerning the $p_T^{}$-dependence of pretzelosity (and other TMDs), 
we remark that in the bag \cite{Avakian:2010br}  and covariant parton 
\cite{Efremov:2009ze} model it is approximately Gaussian which is 
supported by phenomenology \cite{Schweitzer:2010tt}.

%
%

Note that $h_{1T}^{\perp q}(x)$ is not constrained by positivity, 
but $h_{1T}^{\perp(1)q}(x)$ is, see Sec.~\ref{Sec-2:review-pretzelosity}. 
Moreover, $h_{1T}^{\perp q}$ enters cross sections with a prefactor 
of ${\cal O}(p_T^2/M^2)$ from the correlator (\ref{Eq:correlator}).
So in observables effectively the (1)-moment of pretzelosity is relevant, and
the latter is not large in the bag \cite{Avakian:2008dz} or 
light-cone constituent \cite{Pasquini:2008ax} model. In fact, the latter 
predicts \cite{Boffi:2009sh} a small asymmetry (\ref{Eq:AUTpretzel}).
 
The covariant parton model \cite{Efremov:2009ze} predicts  large 
$h_{1T}^{\perp(1) q}(x)$, see Fig.~\ref{Fig-2:TMDs-model}b, and hence
a sizable pretzelosity asymmetry (\ref{Eq:AUTpretzel}) shown in 
Fig.~\ref{Fig-2:TMDs-model}c for the kinematics of the 
CLAS experiment with 12 GeV upgrade \cite{Avakian:2008dz,Efremov:2009ze}.
The predictions are consistent with preliminary SIDIS data 
\cite{Kotzinian:2007uv} showing a zero within error bars effect.
Recent COMPASS data  \cite{Alekseev:2010dm} show a small but non-vanishing
effect and confirm the signs predicted for $h_{1T}^{\perp q}$,
as was discussed in Sec.~\ref{Sec-2:review-pretzelosity}.

One has to bear in mind that so far the exciting relation 
(\ref{Eq:OAM-pretzel}) between TMDs and orbital angular momentum is 
established only in quark models. An important question is: 
to what extent can we trust such models?
One way to address this question consists in reproducing 
(SI)DIS spin observables within a given quark model. 
On the basis of the comparison of model results and data it was found
\cite{Boffi:2009sh} that (light-cone constituent) quark models work 
in the valence $x$-region $0.2 \lesssim x \lesssim 0.6$ 
with an accuracy of (10--30)$\,\%$.

In this context we recall that the absence of gauge degrees of freedom 
implies in quark models certain relations among TMDs (called ``LIRs''),
which hold approximately in QCD upon the neglect of quark-gluon-quark
correlators and current quark mass terms \cite{Metz:2008ib}.
For the collinear twist-3 parton distribution function $g_T^q(x)$
such an approximation works reasonably well \cite{Accardi:2009au}
but it needs to be tested for other TMDs \cite{Kotzinian:2006dw}.
If LIRs were confirmed to be reasonably good approximations,
this would be a necessary (not sufficient) condition for
quark model predictions of the type
(\ref{Eq:pretzel-relation},~\ref{Eq:OAM-pretzel}) to work 
similarly.

\section{Conclusions}
\label{Sec-5:conclusions}

The concept of quark orbital angular momentum is difficult 
to address rigorously in gauge field theories. GPDs and TMDs,
which describe complementary aspects of the nucleon structure, 
give rise to a dual (in quark models equivalent)
picture of quark orbital angular momentum as follows
\begin{eqnarray}
     L_z^q  \;\;\; &\stackrel{\rm QCD + models}{=}&  \;\;\;
     \int{\rm d} x\int{\rm d}^2 b \;\,{\rm GPDs}(x,b) 
     \;\;\;\stackrel{\mbox{\Large\bf !?}}{=} \label{Eq:OAM-GPD}\\
     L_z^q  \;\;\; &\stackrel{\rm quark \; models}{=} &  \;\;\;
     \int{\rm d} x\int{\rm d}^2 p_T^{} {\rm TMDs}(x,p_T^{}) \label{Eq:OAM-TMD}
\end{eqnarray}
where GPDs $= xH^q+xE^q-\widetilde{H}^q$ and TMDs = $- \,h_{1T}^{\perp(1)q}$.
The first relation (\ref{Eq:OAM-GPD}) holds exactly in QCD and in 
consistent models \cite{Ji:1996ek}. The second relation 
(\ref{Eq:OAM-TMD}) holds in a large class of relativistic quark models
\cite{She:2009jq,Avakian:2010br,in-preparation}.
%
%
Quark models catch important features of QCD, and could provide useful 
insights also in the context of GPDs, TMDs and orbital angular momentum, 
provided one uses them responsibly within their range of applicability.
Recent advances \cite{Hagler:2009ni} allow us to test these 
relations in lattice QCD, where due the presently often practioned
omission of disconnected diagrams valence quarks are probed.

\section*{Note added}

In the discussion following the talk it was pointed out that in 
[\refcite{Thomas:2008ga}] the predictions from quark models were 
shown to be compatible with phenomenology and lattice data, resolving 
the ``spin crisis'' from quark model point of view.

\section*{Acknowledgments}

P.~S.~is grateful to the organizers of the 
``4th Workshop on Exclusive Reactions at High Momentum Transfer,''
18-21 May 2010, Jefferson Lab, where this work was reported.
A.~E.\ and O.~T.\ are supported by the Grants RFBR 09-02-01149 and 07-02-91557, 
RF MSE RNP.2.2.2.2.6546 (MIREA) and by the Heisenberg-Landau Program of JINR.
P.~Z.\ is supported by the project AV0Z10100502 of the Academy of Sciences 
of the Czech Republic.
The work was supported in part by DOE contract DE-AC05-06OR23177, under
which Jefferson Science Associates, LLC,  operates the Jefferson Lab.


\end{document}